\documentclass[12pt, english, prb, reprint]{revtex4-1}

\usepackage[T1]{fontenc}
\usepackage[latin9]{inputenc}
\usepackage{refstyle}
\usepackage{amsmath}
\usepackage{graphicx}
\usepackage{wasysym}
\usepackage[colorlinks=true, allcolors=blue]{hyperref}
\usepackage{threeparttable}
\usepackage{attachfile}

\makeatletter

\AtBeginDocument{\providecommand\eqref[1]{\ref{eq:#1}}}
\AtBeginDocument{\providecommand\secref[1]{\ref{sec:#1}}}
\AtBeginDocument{\providecommand\figref[1]{\ref{fig:#1}}}
\AtBeginDocument{\providecommand\tabref[1]{\ref{tab:#1}}}
\providecommand{\tabularnewline}{\\}
\RS@ifundefined{subsecref}
  {\newref{subsec}{name = \RSsectxt}}
  {}
\RS@ifundefined{thmref}
  {\def\RSthmtxt{theorem~}\newref{thm}{name = \RSthmtxt}}
  {}
\RS@ifundefined{lemref}
  {\def\RSlemtxt{lemma~}\newref{lem}{name = \RSlemtxt}}
  {}

\usepackage[version=3]{mhchem}
\usepackage{url}
\usepackage{notes2bib}

\@ifundefined{showcaptionsetup}{}{%
 \PassOptionsToPackage{caption=false}{subfig}}
\usepackage{subfig}
\makeatother

\usepackage{babel}

\begin{document} 
\title{Meta-local density functionals: a new rung on Jacob's ladder}

\author{Susi Lehtola}

\email{susi.lehtola@alumni.helsinki.fi}

\affiliation{Department of Chemistry, University of Helsinki, P.O. Box 55 (A. I. Virtasen
  aukio 1), FI-00014 University of Helsinki, Finland}

\author{Miguel A. L. Marques}

\affiliation{Institut f\"ur Physik, Martin-Luther-Universit\"at
  Halle-Wittenberg, 06120 Halle (Saale), Germany}

\newcommand*\ie{{\em i.e.}}
\newcommand*\eg{{\em e.g.}}
\newcommand*\etal{{\em et al.}}
\newcommand*\citeref[1]{ref. \citenum{#1}}
\newcommand*\citerefs[1]{refs. \citenum{#1}}

\newcommand*\Erkale{{\sc Erkale}}
\newcommand*\LibXC{{\sc LibXC}}
\newcommand*\PySCF{{\sc PySCF}}
\newcommand*\PsiFour{{\sc Psi4}}

\begin{abstract}
The homogeneous electron gas (HEG) is a key ingredient in the
construction of most exchange-correlation functionals of
density-functional theory. Often, the energy of the HEG is
parameterized as a function of its spin density $n_\sigma$, leading to
the local density approximation (LDA) for inhomogeneous
systems. However, the connection between the electron density and
kinetic energy density of the HEG can be used to generalize the LDA by
evaluating it on a geometric average $n_\sigma^\text{avg} ({\bf r}) =
n_\sigma^{1-x} ({\bf r}) \tilde{n}_\sigma^{x} ({\bf r})$ of the local
spin density $n_\sigma ({\bf r})$ and the spin density
$\tilde{n}_\sigma ({\bf r})$ of a HEG that has the local kinetic
energy density $\tau_\sigma ({\bf r})$ of the inhomogeneous
system. This leads to a new family of functionals that we term
meta-local density approximations (meta-LDAs), which are still exact
for the HEG, which are derived only from properties of the HEG, and
which form a new rung of Jacob's ladder of density functionals [AIP
  Conf. Proc. 577, 1 (2001)]. The first functional of this ladder, the
local $\tau$ approximation (LTA) of Ernzerhof and Scuseria
[J. Chem. Phys.  111, 911 (1999)] that corresponds to $x=1$ is
unfortunately not stable enough to be used in self-consistent field
calculations, because it leads to divergent potentials as we show in
this work. However, a geometric averaging of the LDA and LTA densities
with smaller values of $x$ not only leads to numerical stability of
the resulting functional, but also yields more accurate exchange
energies in atomic calculations than the LDA, the LTA, or the tLDA
functional ($x=1/4$) of Eich and Hellgren [J. Chem. Phys. 141, 224107
  (2014)]. We choose $x=0.50$ as it gives the best total energy in
self-consistent exchange-only calculations for the argon
atom. Atomization energy benchmarks confirm that the choice $x=0.50$
also yields improved energetics in combination with correlation
functionals in molecules, almost eliminating the well-known
overbinding of the LDA and reducing its error by two thirds.
\end{abstract}
\maketitle

\section{Introduction \label{sec:Introduction}}

The homogeneous electron gas (HEG) has a special place in the history
of the study of many-electron systems in general, and of
density-functional theory in particular.\citep{Hohenberg1964,
  Kohn1965} In fact, the development of accurate exchange-correlation
functionals typically begins with the local (spin) density
approximation (LDA), whose construction is based on the
exchange-correlation energy of the HEG. This is then modified by an
enhancement factor that depends on the gradient of the density in the
generalized gradient approximation (GGA); the mega-GGA approximation
adds further dependence on the local kinetic energy density and/or the
Laplacian of the electron density.\citep{Becke2014, Jones2015,
  Mardirossian2017a}

LDAs, GGAs, and meta-GGAs form the first three rungs of the so-called
Jacob's ladder of density functional theory,\cite{Perdew2001} each
rung generally leading to approximations of better accuracy.  Although
GGAs and meta-GGAs add more physical information into the density
functional approximation (DFA), they are typically constructed to
maintain exactness for the exchange-correlation energy of the HEG. In
fact, it can be even argued that this is one of the most important
exact conditions that a functional should fulfill.

In this work, we investigate the accuracy of an ansatz, which alike
the LDA is derived from considerations of the HEG only, but which adds
a further dependence on the local kinetic energy density, similarly to
meta-GGAs. These functionals, which we term meta-LDA functionals, thus
constitute a new rung on Jacob's ladder of functionals, that is shown
to fall in accuracy between LDAs and GGAs.

The work is organized as follows. We will describe the theory behind
the meta-LDA approach in \secref{Theory}, and the implementation of
the meta-LDA functionals and the details of our computations in
\secref{Computational-Details}. The accuracy of the novel functionals
is then assessed by benchmarking exchange energies of atoms and
atomization energies of molecules in \secref{Results}. A brief
summary and conclusions are presented in \secref{Conclusions}. Atomic
units are used throughout the manuscript, unless specified otherwise.

\section{Theory \label{sec:Theory}}

The LDA for the exchange energy is derived for the HEG as\citep{Bloch1929, Dirac1930}
\begin{equation}
E_{x}^{\text{LDA}}[n]=-C_{x}\int n^{4/3}(\boldsymbol{r}){\rm d}^{3}r\label{eq:lda}
\end{equation}
where
\begin{equation}
C_{x}=\frac{3}{4}\left(\frac{3}{\pi}\right)^{1/3}.\label{eq:Cx}
\end{equation}
The kinetic energy density of the gas is also known,
\begin{equation}
\tau^{\text{HEG}} = C_F n^{5/3}\label{eq:tau-lda}
\end{equation}
where
\begin{equation}
C_{F}=\frac{3}{10}(3\pi^{2})^{2/3}.\label{eq:Cf}
\end{equation}
Since \eqref{tau-lda} establishes a link between the kinetic energy
density and the electron density, \citet{Ernzerhof1999a} proposed an
exchange functional similar to \eqref{lda} where \eqref{tau-lda} is
used to replace the local density dependence by
\begin{equation}
  \tilde{n}(\boldsymbol{r})=\left[\frac{\tau(\boldsymbol{r})}{C_F}\right]^{3/5}\label{eq:ntau}
\end{equation}
yielding the local $\tau$ approximation (LTA) exchange functional
\begin{align}
E_{x}^{\text{LTA}}[\tau] & =-C_{x}\int\left[\frac{\tau(\boldsymbol{r})}{C_{F}}\right]^{4/5}{\rm d}^{3}r.\label{eq:lta}
\end{align}

Based on the work of \citeauthor{Ernzerhof1999a}, \citet{Eich2014}
suggested another exchange functional where only the energy per unit
particle is written as a function of the fictitious density of
\eqref{ntau}, yielding the tLDA exchange functional
\begin{align}
E_{x}^{\text{tLDA}}[n,\tau] & =-C_{x}\int n(\boldsymbol{r})\tilde{n}^{1/3}(\boldsymbol{r}){\rm d}^{3}r.\label{eq:tlda}
\end{align}

In this work, we show the power of this idea by generalizing the
approach of Ernzerhof, Scuseria, Eich, and Hellgren. We thus replace
the electron density by an effective density $n^\text{eff}({\bf r})$
formed as a weighted combination of the electron density
$n(\boldsymbol{r})$ and the fictitious density computed from
$\tau(\boldsymbol{r})$ as
\begin{equation}
n(\boldsymbol{r})\to
n^{\text{eff}}(\boldsymbol{r})=\tilde{n}^{x}(\boldsymbol{r})n^{1-x}(\boldsymbol{r}).\label{eq:wdens}
\end{equation}
This form interpolates between the LDA ($x=0$), tLDA ($x=1/4$) and LTA
($x=1$) in the case of the exchange functional. Furthermore, it can
also be employed within any LDA correlation functional, allowing us to
generate a complete exchange-correlation ansatz.

We note here that the family of functionals generated by \eqref{wdens}
is actually a member of a general family of functionals that have the
form of an LDA, but which are based on a transformed density variable
\begin{align}
  n(\boldsymbol{r}) \rightarrow n(\boldsymbol{r}) f^\text{mLDA}(t(\boldsymbol{r})),
  \label{eq:densmlda}
\end{align}
where $t(\boldsymbol{r})$ is the (dimensionless) reduced kinetic energy density
\begin{equation}
t(\boldsymbol{r})=\frac{\tau(\boldsymbol{r})}{n^{5/3}(\boldsymbol{r})}.\label{eq:t}
\end{equation}
It is easily seen that LDA functionals operating on a density
transformed according to \eqref{densmlda} are exact for the HEG if the
function $f^\text{mLDA}$ reduces to one for the HEG, i.e.
\begin{equation}
  f^\text{mLDA}(C_F) = 1
  \,.
\end{equation}
Because this procedure generates a meta-GGA-type functional without
gradient dependence from a LDA, we will term these functionals
meta-LDAs.

\section{Computational Details\label{sec:Computational-Details}}

The effective density of \eqref{wdens} can be rewritten in the form of
\eqref{densmlda} as
\begin{align}
f(t) & =\left(\frac{t}{C_F}\right)^{3x/5} \,. \label{eq:n-reduced}
\end{align}
The resulting meta-LDA version of the local exchange functional can be
easily rewritten in terms of an enhancement function
\begin{equation}
  F(t;x) = \left[\left(\frac{t}{C_F}\right)^{3x/5}\right]^{4/3}=\left(\frac{t}{C_F}\right)^{4x/5}.\label{eq:Fx-geom}
\end{equation}
The generalization of the Perdew--Wang 1992 correlation
functional\citep{Perdew1992a} is equally trivial; the density used to
evaluate the energy density is merely re-expressed using
\eqref{wdens}. These new functionals have been implemented in version
5.1.0 of the \textsc{Libxc} library of exchange-correlation
functionals.\citep{Lehtola2018} In \textsc{Libxc}, the derivatives of
the functional are evaluated analytically using the \textsc{Maple}
symbolic algebra program, as is the case for all other functionals in
\textsc{Libxc} as well. Combined with a basis set, these derivatives
can be used to minimize the total energy variationally with respect to
the orbital coefficients within a self-consistent field approach; we
refer to \citeref{Lehtola2020} for discussion.

Fully numerical,\citep{Lehtola2019c} fully variational calculations on
closed and partially closed shell atoms from H to Sr were performed
with the finite element method as implemented in the \textsc{HelFEM}
program,\citep{Lehtola2019a} which allows for an efficient approach to
the complete basis set limit.\citep{Lehtola2019e, Lehtola2020c} The
atomic calculations employed five radial elements, yielding 139
numerical radial basis functions which suffice to converge the energy
to better than $\mu E_h$ precision for these systems.

Molecular calculations on the 183 non-multireference molecules in the
W4-17 dataset\citep{Karton2017} were performed with the \textsc{Psi4}
program.\citep{Smith2020} The \textsc{Psi4} calculations employed the
quadruple-$\zeta$ aug-pcseg-3 basis set,\citep{Jensen2001,
  Jensen2002b, Jensen2014} and a $(100,590)$ quadrature grid. Density
fitting\citep{Sambe1975} was used to accelerate the \textsc{Psi4}
calculations; a universal auxiliary basis set was used for this
purpose.\citep{Weigend2008} 

\section{Results\label{sec:Results}}

\subsection{Atomic calculations\label{subsec:Atomic-calculations}}

\begin{figure*}
\subfloat[Ne]{\begin{centering}
\includegraphics[width=0.33\textwidth]{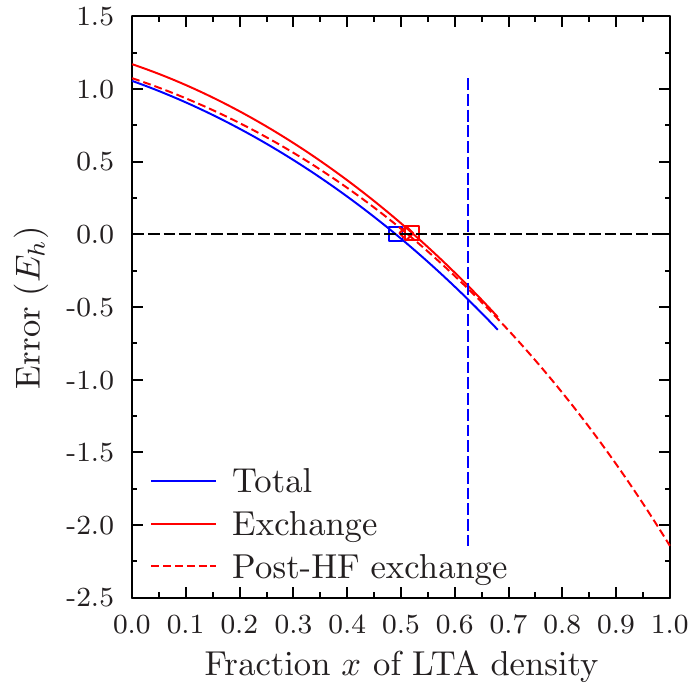}
\par\end{centering}
}\subfloat[Ar]{\begin{centering}
\includegraphics[width=0.33\textwidth]{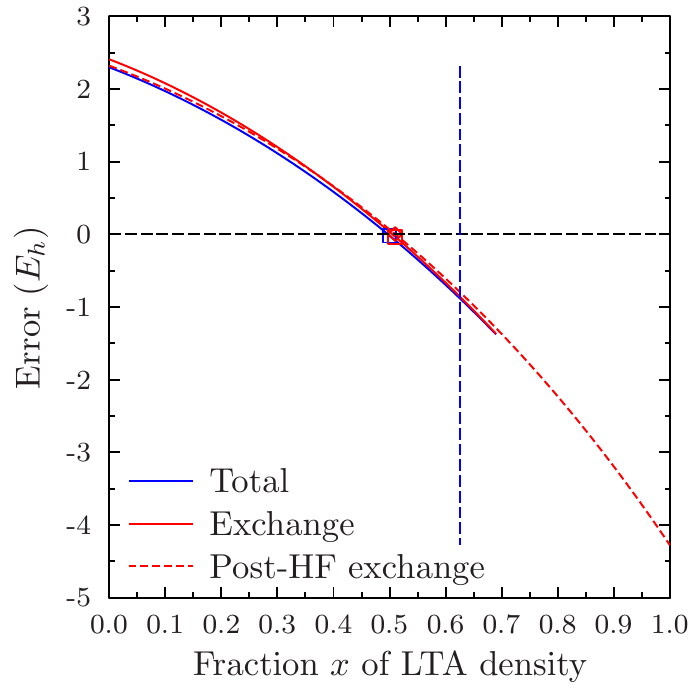}
\par\end{centering}
}\subfloat[Kr]{\begin{centering}
\includegraphics[width=0.33\textwidth]{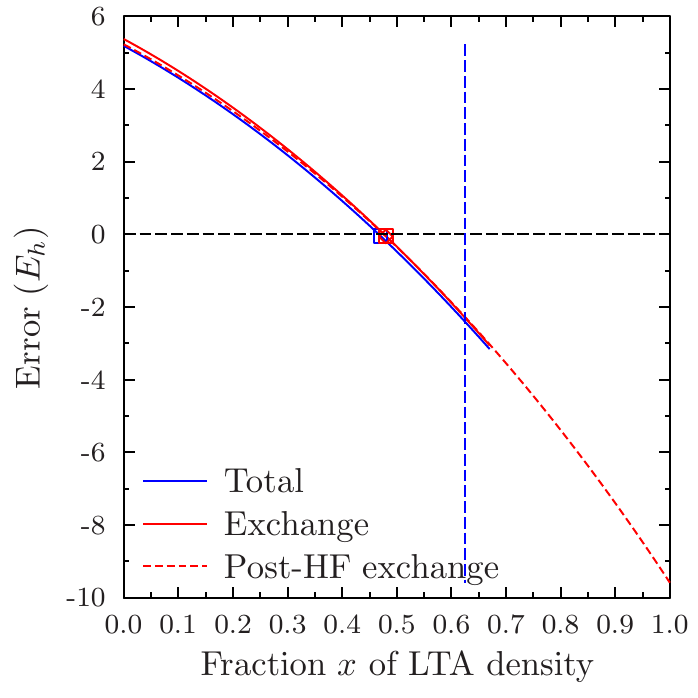}
\par\end{centering}
}

\caption{Errors in self-consistent total (blue solid line) and
  exchange (red solid line) energies of Ne, Ar, and Kr, as well as in
  the perturbative exchange energy calculated on top of the HF density
  (dashed red line). The vertical dashed blue line shows the critical
  value $x=0.625$, see main text. The location of the smallest error
  for the self-consistent total and exchange energies are shown as the
  blue and red squares, respectively, and the one for the perturbative
  exchange energy as red diamonds; however, since the optimal value is
  close to $x=1/2$ for all cases, the markers are on top of each
  other.
\label{fig:noble}}
\end{figure*}

The errors of exchange-only density functional calculations compared
to unrestricted Hartree--Fock (HF) total and exchange energies for
atoms from H to Sr were studied with \textsc{HelFEM}; the reference
unrestricted HF total energies have been recently reported in
\citeref{Lehtola2020c}. Due to the similarity of the results, data is
shown here only for the noble gases Ne, Ar, and Kr in \figref{noble};
the rest of the data can be found in the Supporting Information. In
addition to the self-consistent data, \figref{noble} also shows the
perturbative evaluation of the exchange energy computed on top of the
HF density.

Following \citet{Becke1988a} and \citet{Sun2015} among others, we fit
the parameter $x$ for our meta-LDAs by optimizing the total energy of
the argon atom to the Hartree--Fock reference value, leading to the
choice $x=0.50$. It is noteworthy that in addition to being
quasi-optimal for all systems, $x=0.50$ is also numerically stable for
all the studied atoms. Finally, it also leads to uniformly smaller
errors in the exchange energy than in the LDA and tLDA, which
uniformly underestimate the energy, while LTA grossly overestimates
the energy.

As was already implied above, the self-consistent calculations diverge
for large fractions $x$ of the LTA density. We have analyzed the
instability observed in the calculations; see the Appendix for a
formal analysis. It turns out that the functional form is inherently
unstable for $x>0.625$, since for such values of $x$ the potentials
corresponding to both $n$ and $\tau$ diverge asymptotically to
$-\infty$ for $r \to \infty$. However, it is clear from the results
that the optimal value $x$ for the exchange functional is found at
$x<0.625$.

\subsection{Molecular calculations\label{subsec:Molecular-calculations}}

\begin{table*}
  \caption{Mean absolute error (MAE) and mean error (ME) in
    atomization energies of the non-multireference part of the
    W4-17 test set, computed in the aug-pcseg-3 basis with density
    fitting and a (100,590) grid. The data is divided into
    exchange-only calculations (\ref{tab:xfunc}), and calculations
    including both exchange and correlation (\ref{tab:xcfunc}).
    See the main text for the legend of the functionals shown. To
    clarify the notation, the used values for $x$ in the meta-LDA
    exchange and correlation functionals are also
    shown.
  }
  \begin{centering}
    \subfloat[Results for exchange-only
      calculations. \label{tab:xfunc}]{\begin{centering}
        \begin{threeparttable}
          \begin{tabular}{llr@{\extracolsep{0pt}.}lr@{\extracolsep{0pt}.}l}
            Functional & $x$ & \multicolumn{2}{c}{MAE (kcal/mol)} & \multicolumn{2}{c}{ME (kcal/mol)}\tabularnewline
            \hline
            LDA exchange &  & 28&966 & -12&015\tabularnewline
            \hline
            hLTA exchange$^{a}$ & $1/2$ & 71&235 & -67&512\tabularnewline
            tLTA exchange & $1/3$ & 47&504 & -35&863\tabularnewline
            qLTA exchange$^{b}$ & $1/4$ & 42&181 & -26&070\tabularnewline
            \hline
            HF &  & 144&848 & -144&848\tabularnewline
            B88 exchange &  & 98&177 & -98&177\tabularnewline
            PBE exchange &  & 87&958 & -87&958\tabularnewline
            \hline
            \hline
          \end{tabular}
          \begin{tablenotes}
          \item [$^{a}$] The data for the exchange-only hLTA
            calculation excludes \ce{CH2NH2} for which the SCF
            procedure did not converge.
          \item [$^{b}$] qLTA is the same as the tLDA of
            \citeauthor{Eich2014}.
          \end{tablenotes}
        \end{threeparttable}
        \par\end{centering}
    } \par\end{centering}
    \begin{centering}
      \subfloat[Results for exchange-correlation
        functionals. \label{tab:xcfunc}]{\begin{centering}
          \begin{threeparttable}
            \begin{tabular}{lllr@{\extracolsep{0pt}.}lr@{\extracolsep{0pt}.}l}
            Functional & $x_{\text{exchange}}$ & $x_{\text{correlation}}$ & \multicolumn{2}{c}{MAE (kcal/mol)} & \multicolumn{2}{c}{ME (kcal/mol)}\tabularnewline
            \hline
            LDA-PW92 &  &  & 79&879 & 79&879\tabularnewline
            \hline
            qLTA-qPW92 & $1/4$ & $1/4$ & 61&089 & 60&897\tabularnewline
            tLTA-tPW92 & $1/3$ & $1/3$ & 50&207 & 49&494\tabularnewline
            hLTA-PW92 & $1/2$ & $0$ & 31&388 & 23&913\tabularnewline
            hLTA-hPW92 & $1/2$ & $1/2$ & 26&907 & 14&088\tabularnewline
            \hline
            B88-P86 &  &  & 19&173 & 18&899\tabularnewline
            PBE-PBE &  &  & 18&028 & 17&052\tabularnewline
            TPSS-TPSS &  &  & 12&427 & 11&180\tabularnewline
            B88-LYP &  &  & 8&176 & 1&714\tabularnewline
            \hline
            \hline
          \end{tabular}
        \end{threeparttable}
          \par\end{centering}
      }
      \par\end{centering}
      \label{tab:Errors-in-atomization}
\end{table*}

The application of the functional to atomization energies
\begin{equation}
E^{\text{at}}=\left[\sum_{\text{atoms }A}E(A)\right]-E(\text{molecule})\label{eq:Eat}
\end{equation}
of the non-multireference part of W4-17 yields the errors
\begin{equation}
\Delta E^{\text{at}}=E^{\text{at}}(\text{DFT})-E^{\text{at}}(\text{W4-17})\label{eq:dEat}
\end{equation}
shown in \tabref{Errors-in-atomization}. Due to the higher cost of the
molecular calculations compared to the atomic ones, the new family of
meta-LDA functionals is only studied at select points which suffice
for the present purposes of showing the proof of concept. The points
at which the meta-LDAs are evaluated are indicated by a prefix to the
name of the exchange and correlation functionals: data are presented
for the LDA exchange functional as qLTA (same as
\citeauthor{Eich2014}'s tLDA), tLTA, and hLTA, which stand for for
$x=1/4$, $x=1/3$, and $x=1/2$, respectively. Data is given both for
exchange-only calculations, and for combinations with the Perdew--Wang
(PW92) correlation functional\citep{Perdew1992a} that also admits
meta-LDA generalizations to qPW92, tPW92 and hPW92 for $x=1/4$,
$x=1/3$, and $x=1/2$, respectively.

For comparison, data is also included for the Perdew--Burke--Ernzerhof
exchange-correlation functional;\citep{Perdew1996, Perdew1997}
combinations of the Becke'88 (B88) exchange
functional,\citep{Becke1988a} with the Perdew'86\citep{Perdew1986,
  Perdew1986b} (P86) and Lee--Yang--Parr\citep{Lee1988} (LYP)
correlation functionals; as well as the
Tao--Perdew--Staroverov--Scuseria (TPSS) exchange-correlation
functional.\citep{Tao2003, Perdew2004}

Starting out with the basics, the table demonstrates the well-known
characteristics of HF and LDA: HF severely underbinds molecules due
to the complete neglect of electronic correlation effects, while LDA
overbinds them. Due to the overbinding, exchange-only LDA calculations
are more accurate than those that explicitly include also correlation
contributions, although LDA exchange by itself is slightly underbinding.
In contrast, while the gradient-corrected exchange functionals yield
bad results if used alone, when they are combined with a good gradient-corrected
correlation functional they achieve good accuracy. Jacob's ladder\citep{Perdew2001}
is also visible in the results: more accurate atomization energies
are obtained in the sequence LDA $\to$ PBE $\to$ TPSS.

\begin{figure*}
\subfloat[LDA-PW]{\begin{centering}
\includegraphics[width=0.5\textwidth]{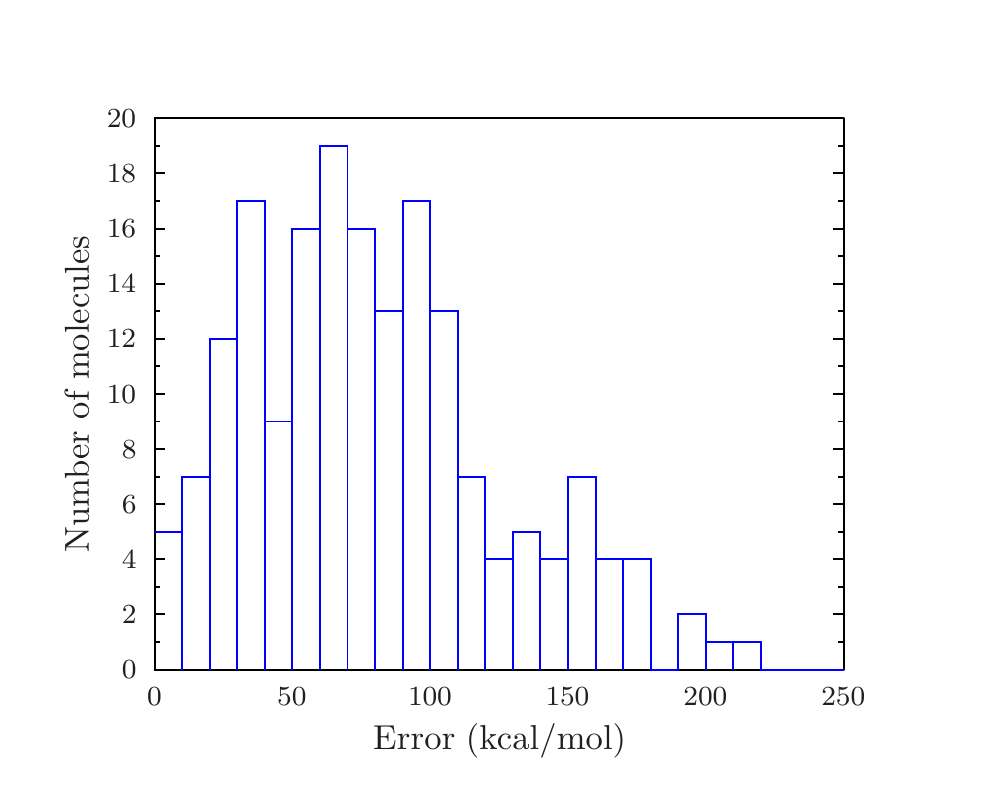}
\par\end{centering}
}\subfloat[hLTA-hPW]{\begin{centering}
\includegraphics[width=0.5\textwidth]{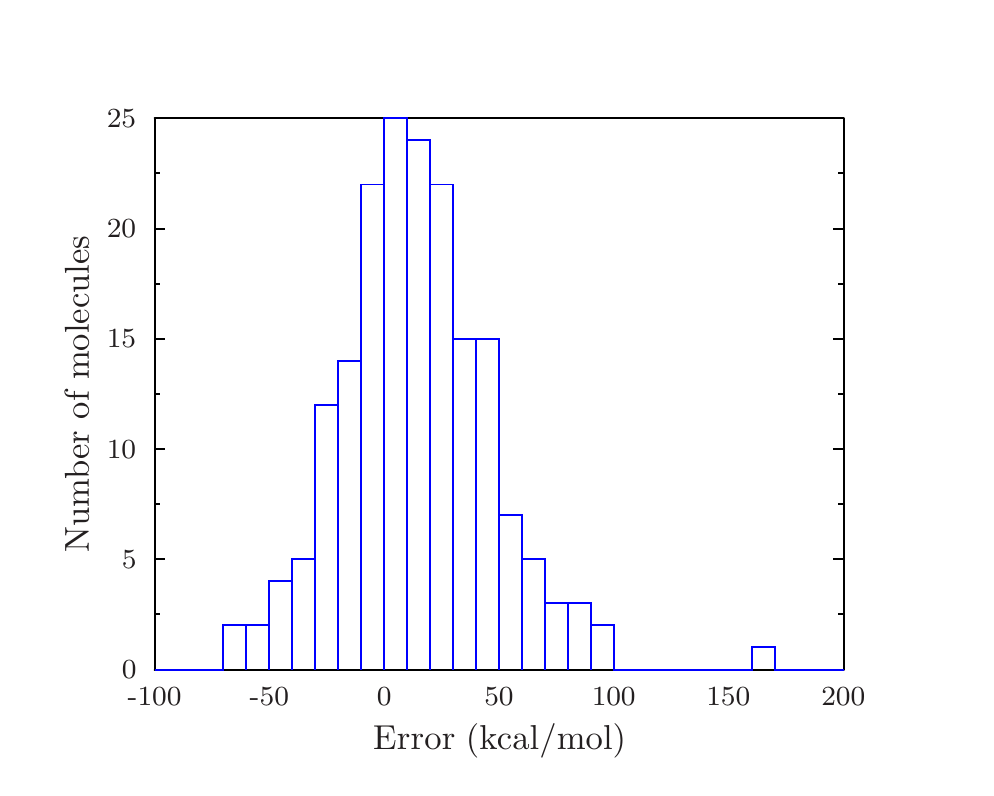}
\par\end{centering}
}

\caption{Error histograms for the atomization energies of the non-multireference
part of W4-17 in the aug-pcseg-3 basis set.\label{fig:Errors-in-atomization}}
\end{figure*}

Also the meta-LDA functionals interestingly show monotonic behavior.
Going from LDA to qLTA to tLTA and, finally, hLTA in exchange-only
calculations leads to systematically increasing underbinding. The
same effect holds also in the presence of correlation: while LDA-PW92
is greatly overbinding, as was already established above, the overbinding
decreases systematically in the sequence LDA-PW92 $\to$ qLTA-qPW92
$\to$ tLTA-tPW92 $\to$ hLTA-hPW92. Like in the case of the atomic
exchange energies, the half-and-half $x=1/2$ mixture of the electron
density with the $\tau$-based density as in the hLTA-hPW92 functional
yields the best results with a mean absolute error almost three times
smaller than in the original LDA-PW92 calculation. This finding is
underlined by the error histograms shown in \figref{Errors-in-atomization}:
while LDA-PW is consistently overbinding, the errors for hLTA-hPW
are almost symmetric, even though the error scale is still large compared
to established GGA functionals.

\section{Summary and Conclusions \label{sec:Conclusions}}

We have proposed a new class of functionals as generalizations of the
established class of local density approximations (LDAs) by including
a fraction $x$ of fictitious density computed from the local kinetic
energy density via a relation derived for the homogeneous electron gas
(HEG). The resulting so-called meta-LDA functionals maintain the
exactness of LDA for the HEG, and are derived from HEG data only
(with the exception of the one parameter $x$ that is fitted to the total
exchange-only energy of the argon atom), but afford much improved
accuracy for inhomogeneous systems, thus forming a new rung on Jacob's
ladder of density functionals in-between LDAs and GGAs. Benchmarks on
both perturbative and self-consistent atomic exchange energies, as
well as molecular atomization energies in the presence of a
correlation functional showed that the half-and-half ratio $x=1/2$
yields quasi-optimal results for both atoms and molecules, almost
fully eliminating the overbinding of LDA and reducing the mean
absolute error in the atomization energies to a third of the original.

Meta-LDAs could also been seen as a better starting point for the
inclusion of an extra dependency in the gradient of the density (as in
a standard GGA), and in the Laplacian of the density and the kinetic
energy density (as in a standard meta-GGA). Due to the extra
flexibility we can expect that these will be better performing than
the parent functionals. For example, the new degree of freedom
introduced with the meta-LDAs could play an important role for, e.g.,
semi-empirical functionals fitted to experimental data. In many of
these cases (see for example on \citerefs{Becke1997, Boese2000,
  Verma2017}), the functionals do not reduce to the LDA for
homogeneous densities, as this would compromise the accuracy of the
functional for other systems. By replacing the standard LDA by a
meta-LDA form in full or in part on could, in principle, obey the
exact condition without compromising accuracy, and at the same time
increase the transferability of the functionals to solids. Of course,
the GGA or meta-GGA enhancement functionals has to be redesigned (or
at least re-optimized) to take the new form into account. Work along
these lines has already started.

\section*{Appendix: instability of the local tau approximation}
The Kohn--Sham electron density is known to behave asymptotically as
$n({\bf r}) \propto \exp(-2\sqrt{-2 \epsilon_\text{HOMO}}r)$ due to
the highest occupied molecular orbital (HOMO) which behaves as
$\psi_\text{HOMO} \propto
\exp(-\sqrt{-2\epsilon_\text{HOMO}}r)$.\cite{Katriel1980} For
simplicity, we will study hydrogenic orbitals of the form
\begin{equation}
  \psi_\sigma ({\bf r}) = \frac {2 \zeta^{3/2} \exp(-\zeta r)}
      {\sqrt{4\pi}} \label{eq:asymptotic}
\end{equation}
to show that the exponentially decaying asymptotic region leads to
problems for $r \to \infty$ for the local tau exchange functional. The
electron density of the wave function in \eqref{asymptotic} is
\begin{equation}
  n_\sigma ({\bf r}) = \frac {\zeta^{3} \exp(- 2\zeta r)}
  {\pi}, \label{eq:den}
\end{equation}
while the kinetic energy density is
\begin{equation}
  \tau_\sigma ({\bf r}) = \frac 1 2 \sum_i |\nabla \psi({\bf r})|^2 =
  \frac {\zeta^5 \exp(-2\zeta r)}{2 \pi}. \label{eq:tau}
\end{equation}
The self-consistent implementation of the meta-LDAs is based on the
potentials $v_n^\sigma$ and $v_\tau^\sigma$, which are defined as the
derivatives of the exchange energy density arising from the
substitution of \eqref{wdens} into \eqref{lda} with respect to
$n_\sigma$ and $\tau_\sigma$, respectively.\cite{Lehtola2020} It is
easy to show using e.g. computer algebra (we used \textsc{Maple 2020}
to obtain these results) that when evaluated on an electron density
and kinetic energy density of the form of \eqref{den, tau}, both of
the potentials $v_n^\sigma$ and $v_\tau^\sigma$ contain a factor of
the form $\exp\left[ (16x-10)r\zeta/15 \right]$, which diverges in the
limit $r\to \infty$ whenever $x > x_\text{crit}$ with the critical
value $x_\text{crit} = 5/8 = 0.625$.

Interestingly, also the choice of a HOMO with a Gaussian form
$\psi_\sigma \propto \exp(-\zeta r^2)$ leads to divergent
potentials---only now of a stronger kind $\exp\left[
  (16x-10)r^2\zeta/15 \right]$---and yields the same critical value
$x_\text{crit}=5/8$. In fact, it can be shown that \emph{all}
asymptotic wave functions of the kind $\psi_\sigma \propto \exp(-\zeta
r^p)$ with $p>0$ lead to divergences of the kind $\exp\left[ (16x-10)
  \zeta r^p /15 \right]$ in $v_n^\sigma$ and $v_\tau^\sigma$. The
total exchange energy, however, is finite in each case.

For $x>x_\text{crit}$ one then has $v_n^\sigma \to -\infty$ and
$v_\tau^\sigma \to -\infty$ for $r \to \infty$, because the potentials
are negative everywhere (as expected for an exchange functional). This
divergence causes convergence problems. Assuming an orthonormal
basis set $\{\chi_\mu\}$, the potentials $v_n^\sigma$ and
$v_\tau^\sigma$ contribute to the Kohn--Sham--Fock matrix
as\cite{Lehtola2020}
\begin{widetext} 
\begin{equation}
  F^\sigma_{\mu \nu} \propto \int \left[ v_n^\sigma ({\bf r}) \chi_\mu
    ({\bf r}) \chi_\nu ({\bf r}) + \frac 1 2 v_\tau^\sigma ({\bf r}) \left[
    \nabla \chi_\mu ({\bf r}) \cdot \nabla \chi_\nu ({\bf r}) \right]
    \right] {\rm d}^3r. \label{eq:Ftau}
\end{equation}
\end{widetext} 
The tentative physical interpretation of the divergent negative
potentials is that displacing electron density toward $r \to \infty$
would lead to a decrease in the energy. Now, if a Gaussian-type or
Slater-type orbital basis set is employed, $\chi_\mu$ and its gradient
will decay asymptotically as $\exp(-\alpha_\mu r^2)$ or
$\exp(-\zeta_\mu r)$, respectively, where $\alpha_\mu$ and $\zeta_\mu$
are the Gaussian and Slater-type exponent, with analogous expressions
for $\nabla \chi_\nu$. Evaluating \eqref{Ftau} then requires
quadrature of an expression with an exponentially decaying part and an
exponentially increasing part, which is numerically unstable, as the
resulting value may be either small or large. The finite element
calculations with \textsc{HelFEM}, in turn, feature localized basis
functions also at large values of $r$. This leads to exponentially
increasing elements of the Kohn--Sham--Fock matrix, making the
self-consistent field algorithm unstable.

In contrast, the potentials arising in the asymptotic region for
$x<0.625$ decay exponentially (like they do in the local density
approximation), making self-consisistent field calculations stable.

\section*{Acknowledgments}
This work has been supported by the Academy of Finland (Suomen
Akatemia) through project number 311149. Computational resources
provided by CSC -- It Center for Science Ltd (Espoo, Finland) and the
Finnish Grid and Cloud Infrastructure (persistent identifier
urn:nbn:fi:research-infras\-2016072533) are gratefully acknowledged.

\section*{Data Availability}
The data that supports the findings of this study are available within
the article and its supplementary material.

\section*{Supporting Information}

\begin{figure*}
\subfloat[Be]{\begin{centering}
\includegraphics[width=0.33\textwidth]{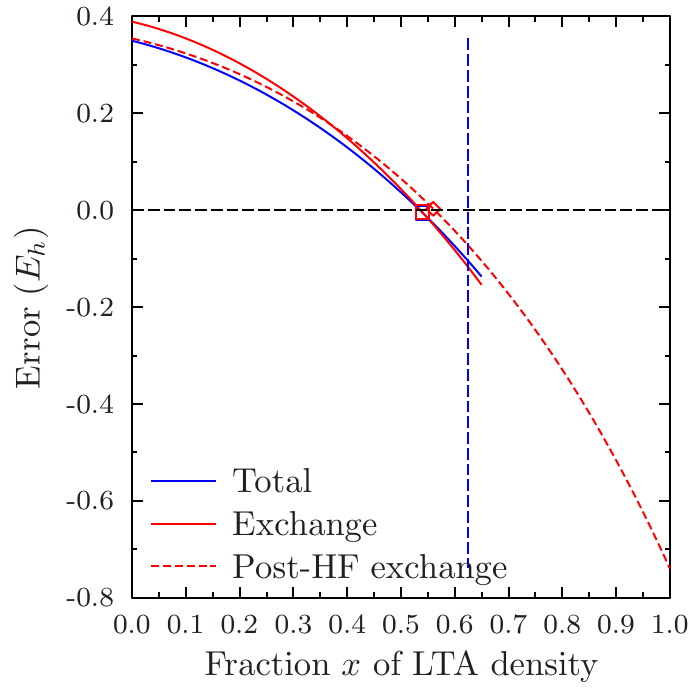}
\par\end{centering}
}\subfloat[Mg]{\begin{centering}
\includegraphics[width=0.33\textwidth]{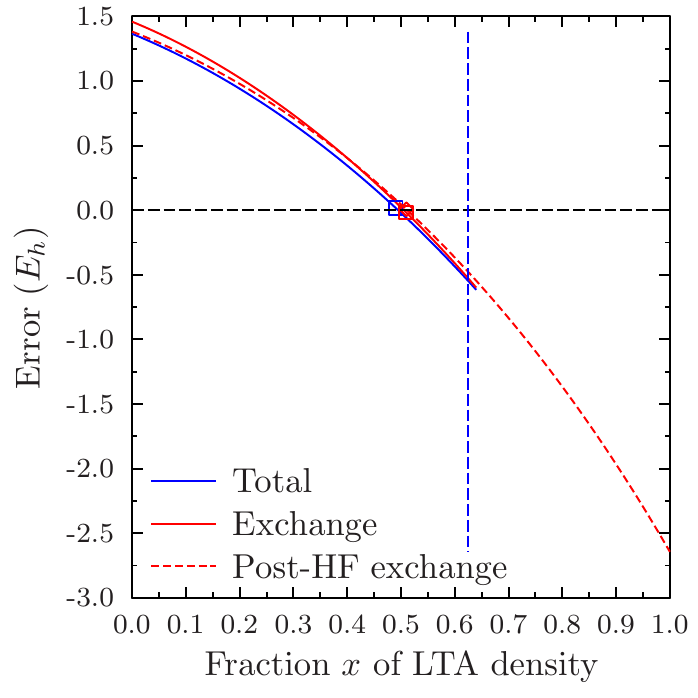}
\par\end{centering}
}\subfloat[Ar]{\begin{centering}
\includegraphics[width=0.33\textwidth]{figs//Ar}
\par\end{centering}
}

\subfloat[Ca]{\begin{centering}
\includegraphics[width=0.33\textwidth]{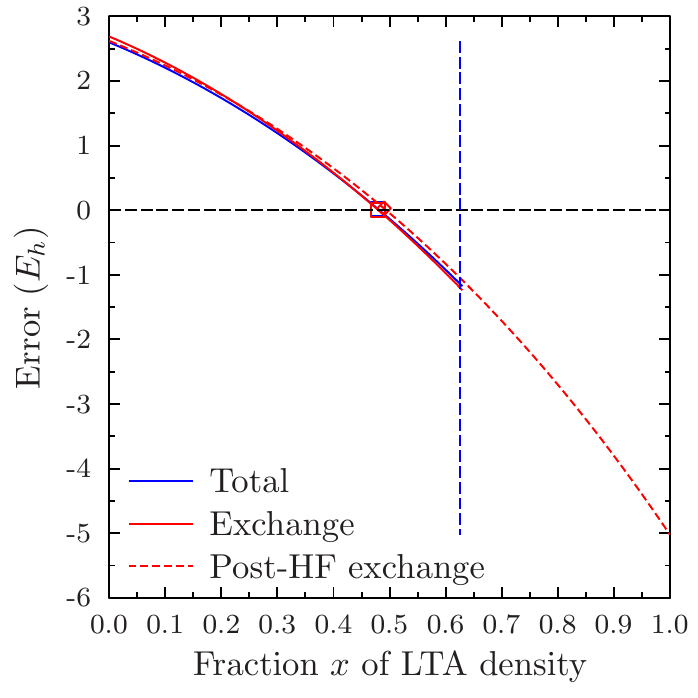}
\par\end{centering}
}\subfloat[Ni]{\begin{centering}
\includegraphics[width=0.33\textwidth]{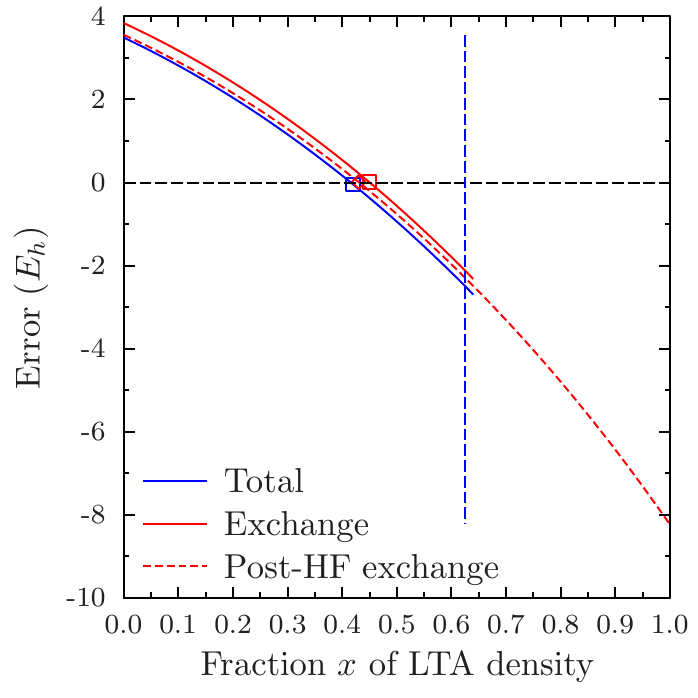}
\par\end{centering}
}\subfloat[Zn]{\begin{centering}
\includegraphics[width=0.33\textwidth]{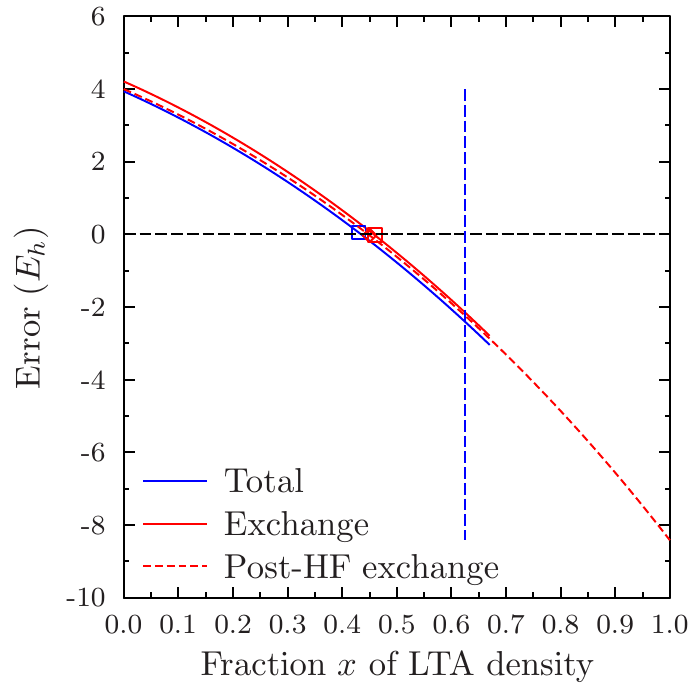}
\par\end{centering}
}
\caption{Errors in self-consistent total (blue solid line) and exchange (red
solid line) energies of closed-shell atoms, as well as in the perturbative
exchange energy calculated on top of the HF density (dashed red line).
The location of the smallest error for the self-consistent total and
exchange energies are shown as the blue and red squares, respectively,
and the one for the perturbative exchange energy as red diamonds.
\label{fig:closedshell}}

\end{figure*}

\begin{figure*}
\subfloat[H]{\begin{centering}
\includegraphics[width=0.33\textwidth]{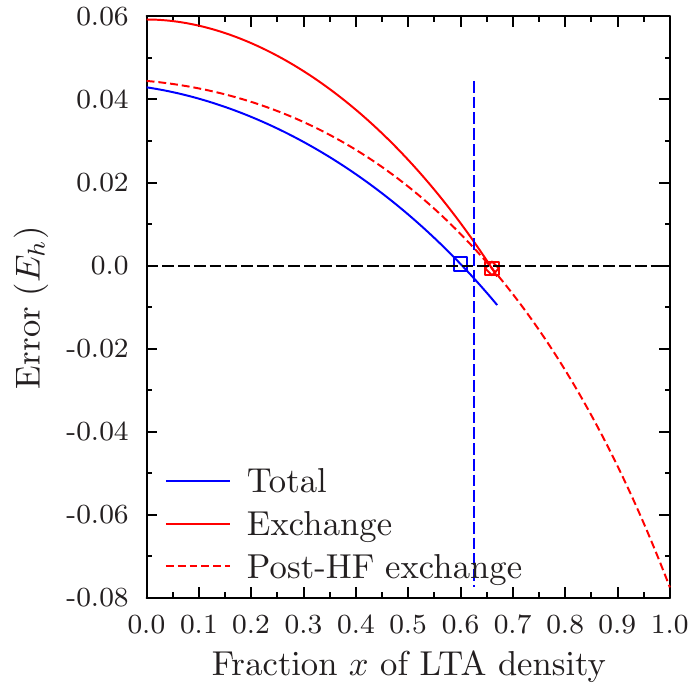}
\par\end{centering}
}\subfloat[Li]{\begin{centering}
\includegraphics[width=0.33\textwidth]{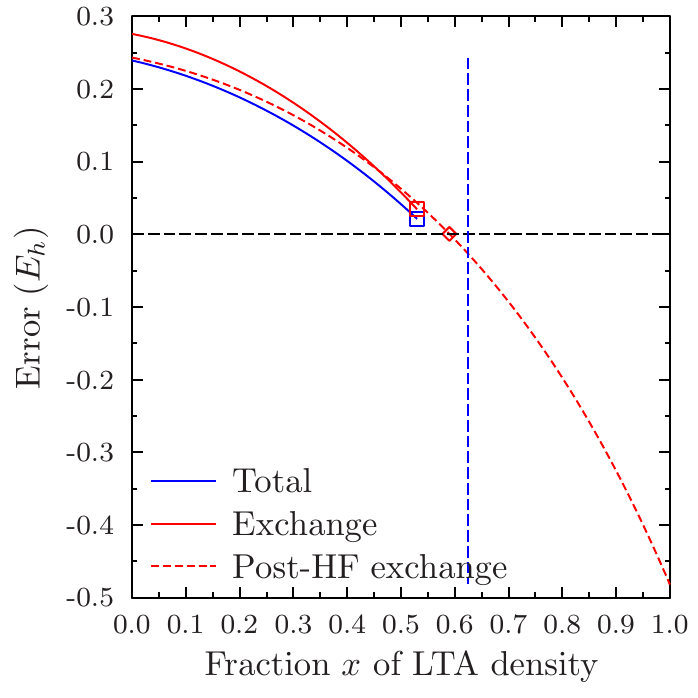}
\par\end{centering}
}\subfloat[C]{\begin{centering}
\includegraphics[width=0.33\textwidth]{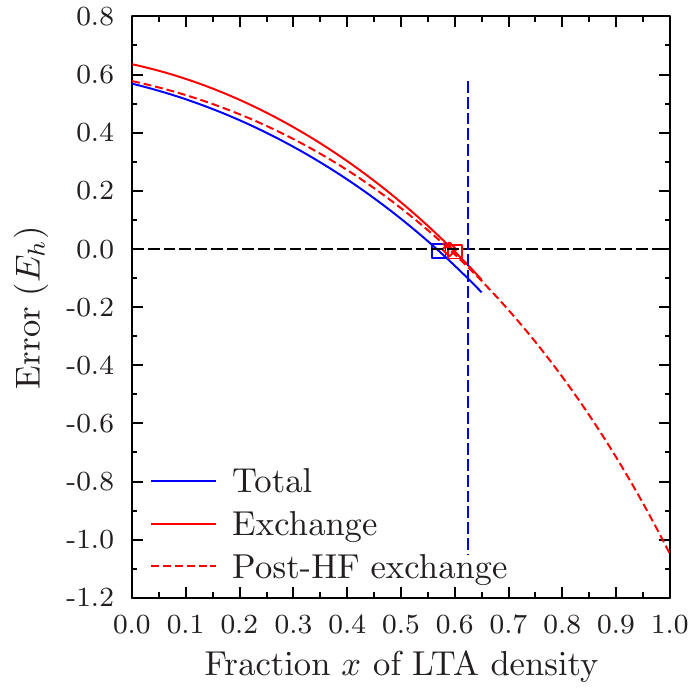}
\par\end{centering}
}

\subfloat[N]{\begin{centering}
\includegraphics[width=0.33\textwidth]{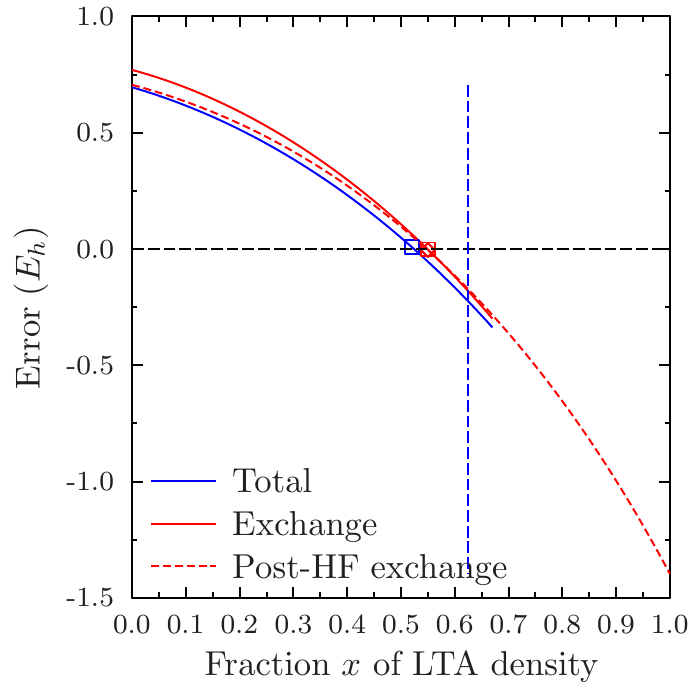}
\par\end{centering}
}\subfloat[Na]{\begin{centering}
\includegraphics[width=0.33\textwidth]{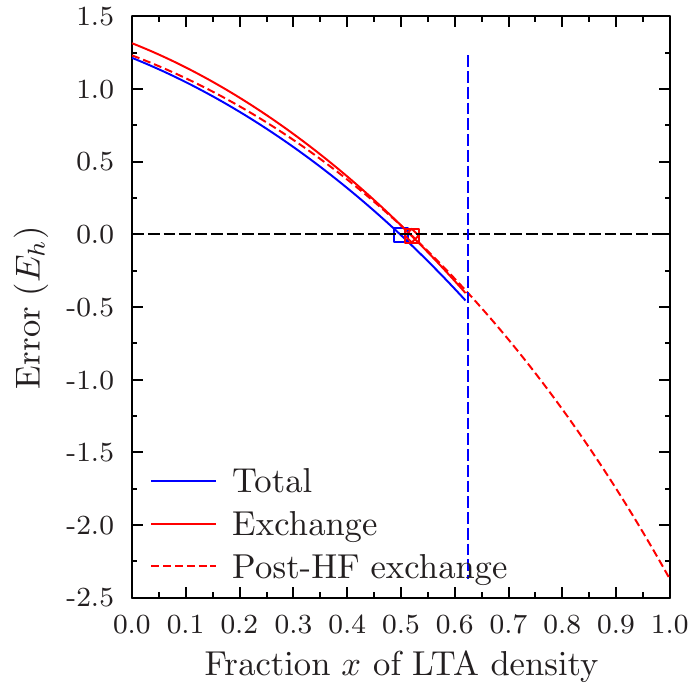}
\par\end{centering}
}\subfloat[Si]{\begin{centering}
\includegraphics[width=0.33\textwidth]{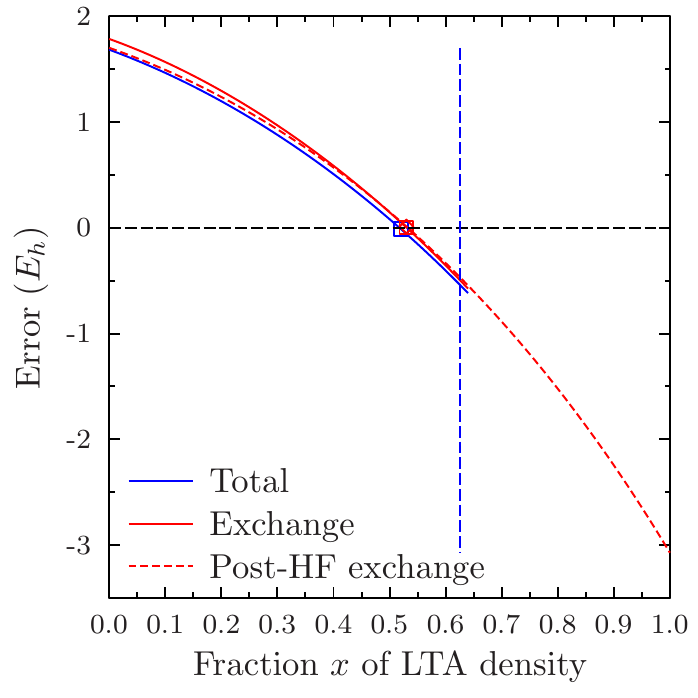}
\par\end{centering}
}

\caption{Errors in self-consistent total (blue solid line) and exchange (red
solid line) energies of partially closed-shell atoms, as well as in
the perturbative exchange energy calculated on top of the HF density
(dashed red line). The location of the smallest error for the self-consistent
total and exchange energies are shown as the blue and red squares,
respectively, and the one for the perturbative exchange energy as
red diamonds.\label{fig:openshell}}
\end{figure*}

\begin{figure*}
\subfloat[P]{\begin{centering}
\includegraphics[width=0.33\textwidth]{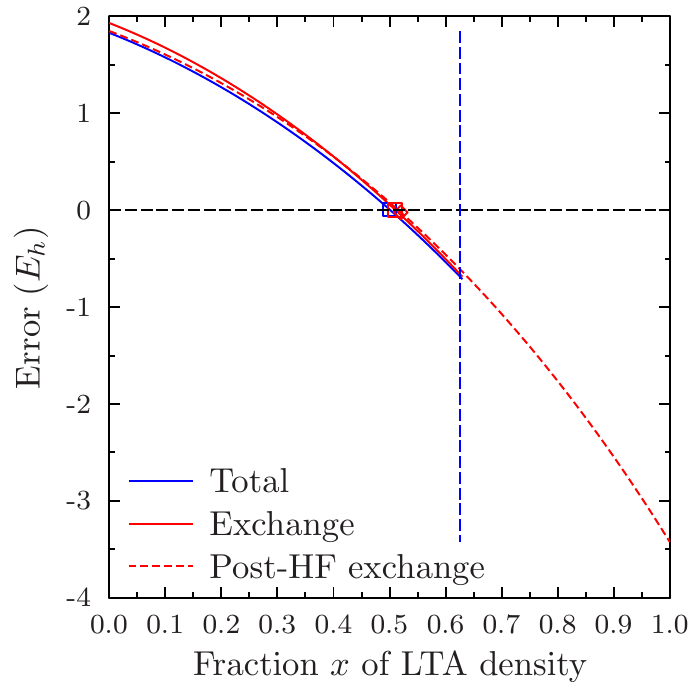}
\par\end{centering}
}\subfloat[K]{\begin{centering}
\includegraphics[width=0.33\textwidth]{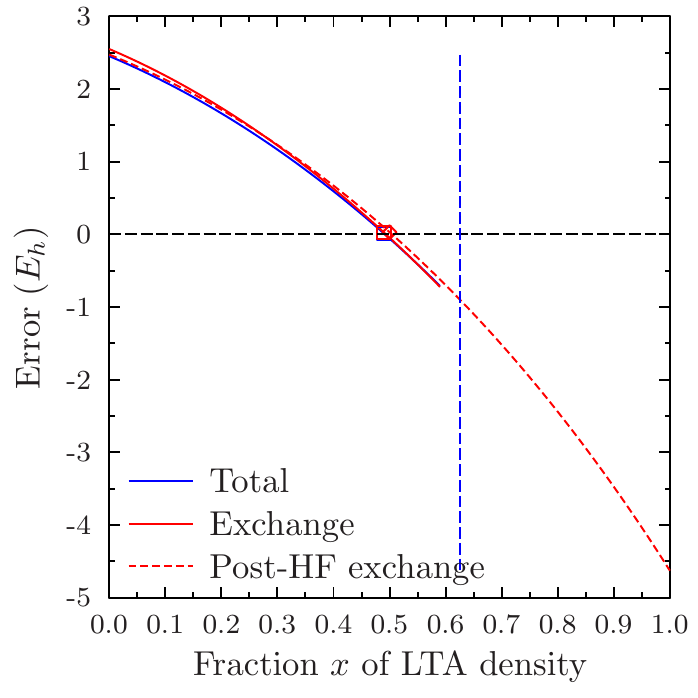}
\par\end{centering}
}\subfloat[V]{\begin{centering}
\includegraphics[width=0.33\textwidth]{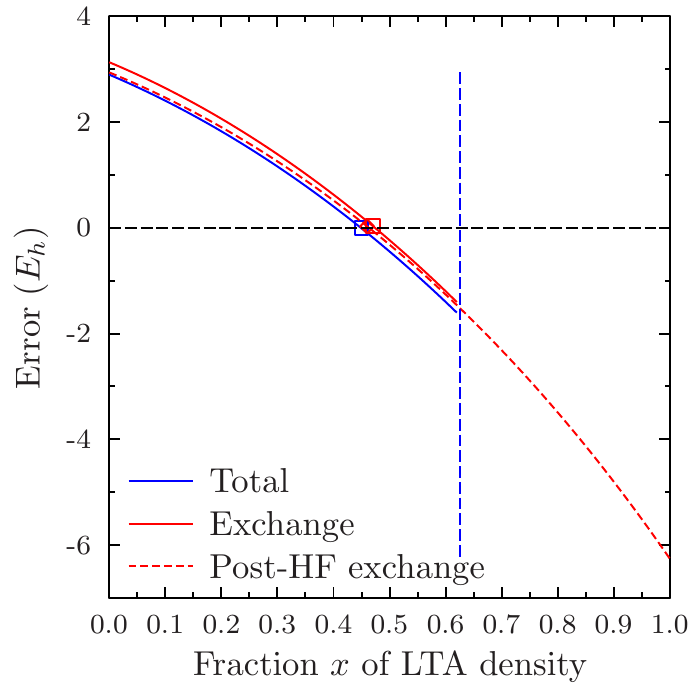}
\par\end{centering}
}

\subfloat[Cr]{\begin{centering}
\includegraphics[width=0.33\textwidth]{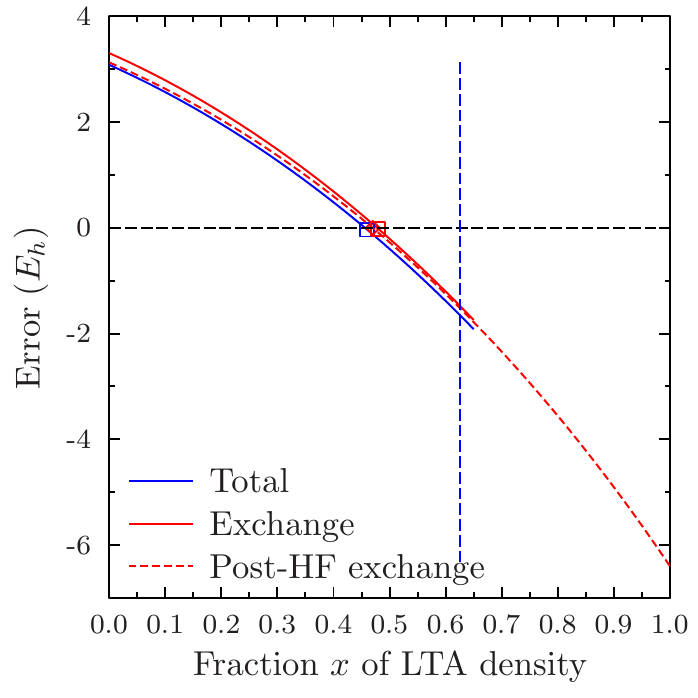}
\par\end{centering}
}\subfloat[Mn]{\begin{centering}
\includegraphics[width=0.33\textwidth]{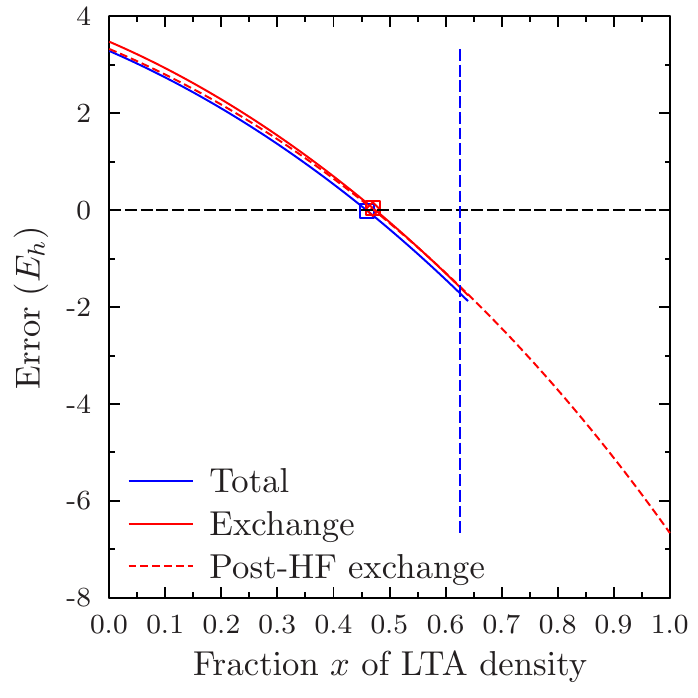}
\par\end{centering}
}\subfloat[Cu]{\begin{centering}
\includegraphics[width=0.33\textwidth]{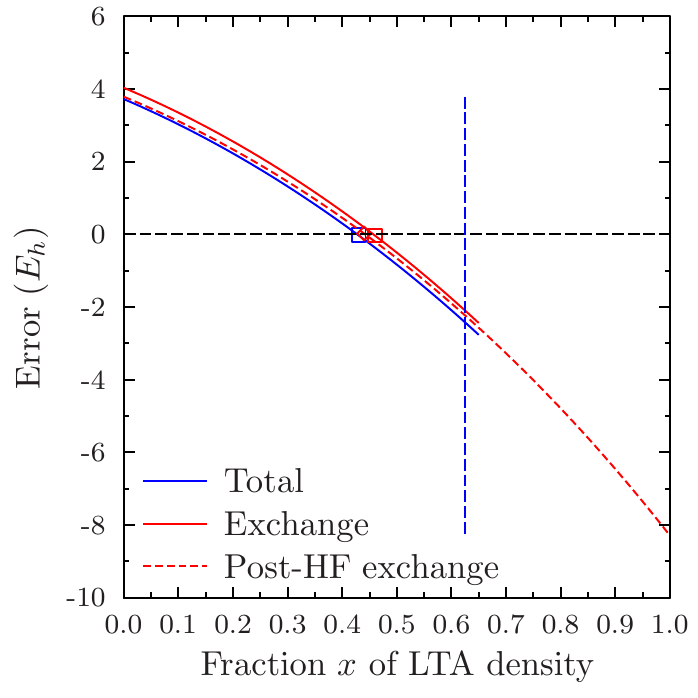}
\par\end{centering}
}

\caption{Errors in self-consistent total (blue solid line) and exchange (red
solid line) energies of partially closed-shell atoms, as well as in
the perturbative exchange energy calculated on top of the HF density
(dashed red line). The location of the smallest error for the self-consistent
total and exchange energies are shown as the blue and red squares,
respectively, and the one for the perturbative exchange energy as
red diamonds.\label{fig:openshell2}}
\end{figure*}

The errors of exchange-only density functional calculations compared
to unrestricted HF total and exchange energies for atoms from H to Sr
are shown in \figref{closedshell} for closed-shell atoms (excluding
Ne, Ar, and Kr that were presented in the main text), and
in \figref{openshell, openshell2} for the partially closed-shell
atoms. In addition to the self-consistent data,
\figrangeref{closedshell}{openshell2} also show a perturbative
evaluation of the exchange energy computed on top of the HF
density. \textattachfile[color=0 0 1]{atomization_energies.txt}{The full list of atomization energies
is attached here in plain text.}

\bibliography{citations}


\end{document}